\pdfoutput=1
\documentclass[prd,twocolumn,superscriptaddress,preprintnumbers,nofootinbib]{revtex4}
\usepackage{graphicx}
\usepackage{epsfig}
\usepackage{bm}
\usepackage{latexsym,amssymb,amsmath,amsfonts,amssymb,txfonts,pxfonts,wasysym,float}
\usepackage{mathrsfs}
\usepackage{color}
\usepackage{lettrine}
\usepackage{lipsum}
\usepackage{enumitem}

\newcommand{\postscript}[2]{\setlength{\epsfxsize}{#2\hsize}
   \centerline{\epsfbox{#1}}}

\usepackage[usenames,dvipsnames]{xcolor}
\definecolor{orange}{cmyk}{0,0.5,1,0}
\definecolor{rossoCP3}{cmyk}{0,.88,.77,.40}
\definecolor{graa}{rgb}{0.8,0.8,0.8}
\definecolor{blaa}{rgb}{0.2,0.2,0.6}

\begin{document}
\preprint{MPP-2020-72}
\preprint{LMU-ASC 17/20}

\title{\color{rossoCP3} Dark energy, Ricci-nonflat spaces, and the Swampland}

\author{Luis A. Anchordoqui}

\affiliation{Department of Physics and Astronomy,  Lehman College, City University of
  New York, NY 10468, USA
}

\affiliation{Department of Physics,
 Graduate Center, City University
  of New York,  NY 10016, USA
}

\affiliation{Department of Astrophysics,
 American Museum of Natural History, NY
 10024, USA
}

\author{Ignatios Antoniadis}
\affiliation{Laboratoire de Physique Th\'eorique et Hautes \'Energies - LPTHE
Sorbonne Universit\'e, CNRS, 4 Place Jussieu, 75005 Paris, France
}

\affiliation{Albert Einstein Center, Institute for Theoretical Physics
University of Bern, Sidlerstrasse 5, CH-3012 Bern, Switzerland
}

\author{Dieter\nolinebreak~L\"ust}

\affiliation{Max--Planck--Institut f\"ur Physik,  
 Werner--Heisenberg--Institut,
80805 M\"unchen, Germany
}

\affiliation{Arnold Sommerfeld Center for Theoretical Physics 
Ludwig-Maximilians-Universit\"at M\"unchen,
80333 M\"unchen, Germany
}

\author{Jorge F. Soriano}

\affiliation{Department of Physics and Astronomy,  Lehman College, City University of
  New York, NY 10468, USA
}

\affiliation{Department of Physics,
 Graduate Center, City University
  of New York,  NY 10016, USA
}

\begin{abstract}
  \vskip 2mm \noindent
It was recently pointed out that the existence of dark energy imposes highly
restrictive constraints on effective
field theories that satisfy the Swampland conjectures. We provide a
critical confrontation of these constraints with the cosmological
framework emerging from the Salam-Sezgin model and its string realization by
Cveti\v c, Gibbons, and Pope. We also discuss the implication of the
constraints for string model building.
\end{abstract}

\maketitle

Very recently, Montefalcone, Steinhardt, and Wesley (MSW) pointed out that
fundamental theories which are based on compactification from extra
dimensions struggle to accommodate a period of accelerated cosmological
expansion~\cite{Montefalcone:2020vlu}. More concretely, they derived constraints on the subset of
``consistent looking'' (3+1) dimensional effective quantum field
theories coupled to gravity that satisfy the Swampland
conjectures~\cite{ArkaniHamed:2006dz,Ooguri:2006in,Klaewer:2016kiy,Ooguri:2018wrx,Grimm:2018ohb,Heidenreich:2018kpg,Ooguri:2016pdq,Palti:2017elp,Obied:2018sgi,Andriot:2018wzk,Cecotti:2018ufg,Garg:2018reu,Klaewer:2018yxi,Heckman:2019bzm,Lust:2019zwm,Bedroya:2019snp,Kehagias:2019akr,Blumenhagen:2019vgj}
(for reviews see~\cite{Brennan:2017rbf,Palti:2019pca}) and
thereby are also consistent with string theory~\cite{Vafa:2005ui}. In a recent study, we
developed a  concrete realization of the cosmological string
framework of fading dark matter~\cite{Agrawal:2019dlm} that can accommodate a period of
accelerated expansion~\cite{Anchordoqui:2019amx}. In this Letter, we confront the predictions of
our model with the constraints derived in~\cite{Montefalcone:2020vlu} and we demonstrate that it
remains a  viable framework to explain the overall data sets of
the latest cosmological observations. 

We begin by summarizing some desirable features of effective field
theories that are inherited from properties of the overarching string
theory.  The Swampland conjectures closely related to our study are
those germane to effective scalar field theories canonically coupled
to gravity and endowed with a canonical kinetic term, which dominates the
energy density of the present epoch universe. For these theories to be
consistent with string theory, the following two conditions are conjectured to hold:
\begin{itemize}[noitemsep,topsep=0pt]
\item Distance Swampland conjecture: If a scalar field
transverses a trans-Planckian range in the moduli space, a tower of string states becomes
light exponentially with increasing distance~\cite{Ooguri:2006in,Klaewer:2016kiy, Ooguri:2018wrx,Grimm:2018ohb,Heidenreich:2018kpg}.
\item de Sitter conjecture: The gradient of the potential $V$
  must satisfy either the lower bound, $M_{\rm Pl}|\nabla V| \geq c
  V$ or else must satisfy
$M_{\rm Pl}^2 {\rm min} (\nabla_i \nabla_jV ) \leq - c'V$, where $c$ and
$c'$ are positive order-one numbers in Planck units and $M_{\rm Pl}$
is the reduced Planck mass~\cite{Obied:2018sgi,Ooguri:2018wrx}.
\end{itemize}
For the purposes of this study, however, we can ignore the criterium 
that restricts near-zero slope because we are considering the specific
application of quintessence scalar fields as models for dark energy.

A {\it key} assumption
in the derivation of the MSW constraints is that the internal space
should be 
compact and conformally Ricci flat, and hence without loss of generality the
metric tensor of the 10-dimensional space can be written as
\begin{equation}
ds_{10} =e^{2\Omega (t,y)}  g_{\mu \nu}^{\rm FRW}(t)  dx^\mu  dx^\nu +
e^{-2\Omega (t,y)}  \bar g_{mn}^{\rm RF}(t,y)   dy^m 
dy^n \,,
\label{msw}
\end{equation}  
where $g^{\rm FRW}$ is the flat Friedmann-Robertson-Walker metric with
time-dependent scale factor $\bar a(t)$, Greek subscripts ($\mu$, $\nu$) are the indices
along the non-compact dimensions with coordinates $x_\mu$, Latin subscripts
($m, n$) are the indices along the 6 compact extra dimensions with
coordinates $y_m$, and the metric of the internal space is chosen
such that $\bar g^{\rm RF}$ has vanishing Ricci scalar curvature with
warp factor $\Omega$~\cite{Wesley:2008fg,Steinhardt:2008nk}. For
compact spaces with
this specific structure, the expansion
rate can be expressed in terms of the 4-dimensional effective scale
factor $a \equiv e^{\chi/2} \bar a$, with
$e^{\chi} \equiv \int e^{2\Omega} \sqrt{g_{10}} \ d^6y$, and the variation
of Newton's constant $G_4$ can be related to the Hubble parameter $H$
according to $\dot G_4/G_4 = - H \kappa$, where
$\kappa = H^{-1} \int e^{2\Omega} \ \varkappa \sqrt{g_6} \ d^6y$,
$g_{mn} \equiv e^{-2\Omega} \bar g_{mn}^{\rm RF}$,
and the time variation of $\varkappa$ drives the local expansion of
the extra dimensions~\cite{Steinhardt:2010ij}. Now, using limits on the instantaneous variation
of $G_4$ today~\cite{Will} MSW derived constraints to be imposed on the
$\kappa (a)$ trajectories for quintessence scalar field dark energy
$\chi$ with potential $V_\chi \propto e^{\lambda \chi}$, where
$\lambda \sim {\cal O}(1)$. It turns out that for $\lambda <1$, the
computed values of $\kappa (a=1)$ are outside the $3\sigma$ range of the
observed instantaneous value of $\dot G_4/G_4$ today~\cite{Montefalcone:2020vlu}.

By all means, the metric of the internal manifold is not always factorable in terms of a warping factor times a Ricci flat
space. A particular string framework where the internal space is not conformally Ricci flat
is that of the Salam-Sezgin model~\cite{Salam:1984cj} with its string
realization by Cveti\v c, Gibbons, and Pope~\cite{Cvetic:2003xr}. The Salam-Sezgin model
is fairly simple, it describes the compactification of a 6-dimensional
supergravity to four dimensions with a monopole background on a
2-sphere, allowing for time dependence of the 6-dimensional dilaton $\phi$ and the breathing mode of the sphere $f$,
while tolerating a 4-dimensional metric with a Friedmann-Robertson-Walker
form~\cite{Anchordoqui:2007sb}.  The metric tensor of the
6-dimensional spacetime is given by
\begin{equation}
ds_6^2 = e^{2f}\, \Big[- dt^2 + e^{2h}d\vec x^{\;2} + r_c^2 \, (d\vartheta^2
+ \sin^2 \vartheta d\varphi^2) \Big],
\end{equation}
where $r_c$ is the compactification radius and $h = \ln \bar a$. The gauge field $F_{\vartheta \varphi}=-b \sin
\vartheta$ is excited on $S^2$ supporting the monopole configuration~\cite{Salam:1984cj}. 

In terms of linear combinations of the $S^2$ moduli field $f = \sqrt{G_4}\, (X-Y)/4$ and
the 6-dimensional dilaton $\phi =\sqrt{G_4}\, (X+Y)/2$, the 4-dimensional effective potential in the Einstein frame
consists of a pure exponential function of a quintessence field $Y$
(which is the 4-dimensional dilaton) times a quadratic polynomial in the
field $e^{-X}$. It turns out that $X$ is a source
of cold dark matter, with a mass proportional to an exponential function of the
quintessence field. When making the volume of the 2-sphere large, namely
for large values of $Y$, there appears a tower of states, which
according to the infinite distance swampland conjecture becomes
exponentially massless. If the standard model fields are confined on
Neveu-Schwarz 5-branes~\cite{Antoniadis:2001sw} the 6-dimensional gauge couplings are
independent of the string dilaton in the string frame, and upon compactification to four
dimensions the 4-dimensional gauge couplings depend on $X$ (rather
than the dilaton $Y$) which is fixed at the minimum of the
potential~\cite{Anchordoqui:2019amx}. This avoids direct couplings of the dilaton to matter
suppressing extra forces competing with gravity. The asymptotic behavior of the Hubble parameter,
$h \sim \ln t$, leads to a conformally flat Friedmann-Robertson-Walker metric
for large times. The de Sitter (vacuum) potential energy density is
characterized by an exponential behavior $V_Y \propto e^{\sqrt{2}
  Y}$. Asymptotically, this represents the crossover situation with
the equation of state for the quintessence field $w_Y = -1/3$, implying expansion at constant velocity with $Y$ varying logarithmically
$Y \sim -\ln t$~\cite{Antoniadis:1988aa,Antoniadis:1988vi}. The deviation from constant velocity expansion into a brief accelerated phase
encompassing the recent past (redshift $z\alt 6$) makes the model
phenomenologically viable~\cite{Anchordoqui:2019amx}.

The Salam-Sezgin model can be uplifted to
 obtain a full  Type I string configuration, where the metric tensor
 takes the form
\begin{widetext}
\begin{equation}
  d s_{10}^2  =  (\cosh 2 \rho)^{1/4} e^{\phi/2} \left\{e^{-\phi} ds_6^2 + 
  dz^2 + \frac{4}{\xi}  \left[d \rho^2 + \frac{\cosh^2 \rho}{\cosh
  2\rho} \left(d \alpha - \sqrt{\frac{\xi}{8}} b
\cos\vartheta d\varphi \right)^2 
 + \frac{\sinh^2 \rho}{\cosh 2\rho} \left(d\beta
+\sqrt{\frac{\xi}{8}} b
\cos\vartheta d\varphi \right)^2  \right] \right\},
\label{cgp}
\end{equation}
\end{widetext}
where $\rho,z,\alpha,\beta$ are the four extra coordinates, $\xi$ is
the rescaled gauge coupling, and the 10-dimensional
dilaton (denoted by $\phiup$) satisfies
\mbox{$e^{\phiup}= e^{-\phi}/\sqrt{\cosh 2
    \rho}$}~\cite{Cvetic:2003xr}. As can be read off by inspection of
(\ref{cgp}) the 6-dimensional metric tensor of the internal space
cannot be factorized to conform with (\ref{msw}), and therefore the
MSW constraint on $\dot G_4/G_4$ can be evaded.

A point worth noting at this juncture is that the uplifted procedure
leading to (\ref{cgp}) implies a non-compact internal manifold. As a consequence, the
string coupling constant, $g_s = e^\phiup$, goes to zero at large distances $\rho$ in the
internal directions. In addition, the ratio 
$G_{10}/G_6 = 16 \pi^2 \xi^{-3/2}  \int dz \int_0^\infty d \rho \sinh 2 \rho$,
points to a vanishing $G_6$ to accommodate the diverging $\rho$ integration. However, the
metric in~(\ref{cgp}) can be interpreted within the context of  a Klebanov-Strassler throat like 
in~\cite{Giddings:2001yu}, with
$0 \leq \rho \leq L,$ 
$L \gg 1$ being an infrared cutoff, 
to obtain a compact internal space and therefore $G_6 \neq 0$. Alternatively, in the spirit of~\cite{Stelle:2020mmg}, we can introduce a warping factor
in the non-compact space to make the $\rho$ integration finite and
obtain
\begin{equation}
G_4 = \frac{243 \ [\zeta(3)]^2 \ G_{10} \ \xi^{5/2}}{16 \ \pi \ \ell_z} \, ,
\end{equation}
where $\int dz = \ell_z$. Now, since the
cosmological parameters determined elsewhere
~\cite{Anchordoqui:2019amx} are independent
of the moduli fields but the one supporting the Salam-Sezgin monopole, we can
always select the time variation of the compact dimension $\ell_z$ to
accommodate the $\dot G_4/G_4$ constraints.

De facto the Salam-Sezgin model is a supersymmetrization of a non-critical
string with exponential tree-level dilaton potential proportional to
the central charge deficit~\cite{Salam:1984cj}. Cveti\v c, Gibbons, and Pope~\cite{Cvetic:2003xr} provided the
10-dimensional compactification on a non-compact
space that we adopted in~\cite{Anchordoqui:2019amx}, but this is
just one example. All one needs is some internal (super-)conformal field
theory (CFT) with the appropriate central charge (bigger than 4 to
account for the `non-compactness' in a $\sigma$-model approach) to
go to six dimensions. In the CFT approach there is no 10-dimensional
Planck constant since the internal central charge is bigger than
4. Instead there is the string scale and a 6-dimensional Planck scale,
and  therefore there is no problem of non compactness.

\begin{figure*}[tb] 
\begin{minipage}[t]{0.49\textwidth}
\postscript{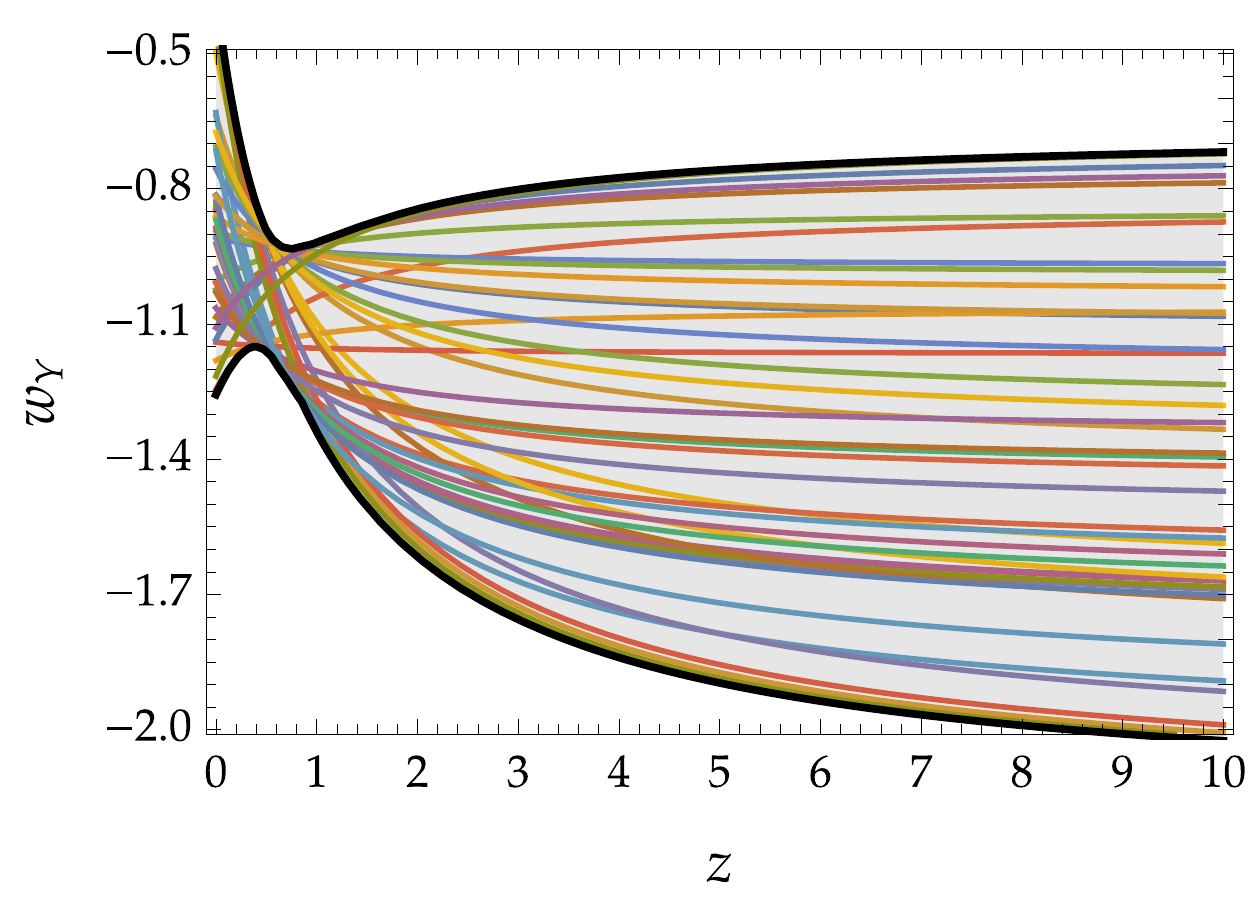}{0.99}
\end{minipage}
\begin{minipage}[t]{0.49\textwidth}
\postscript{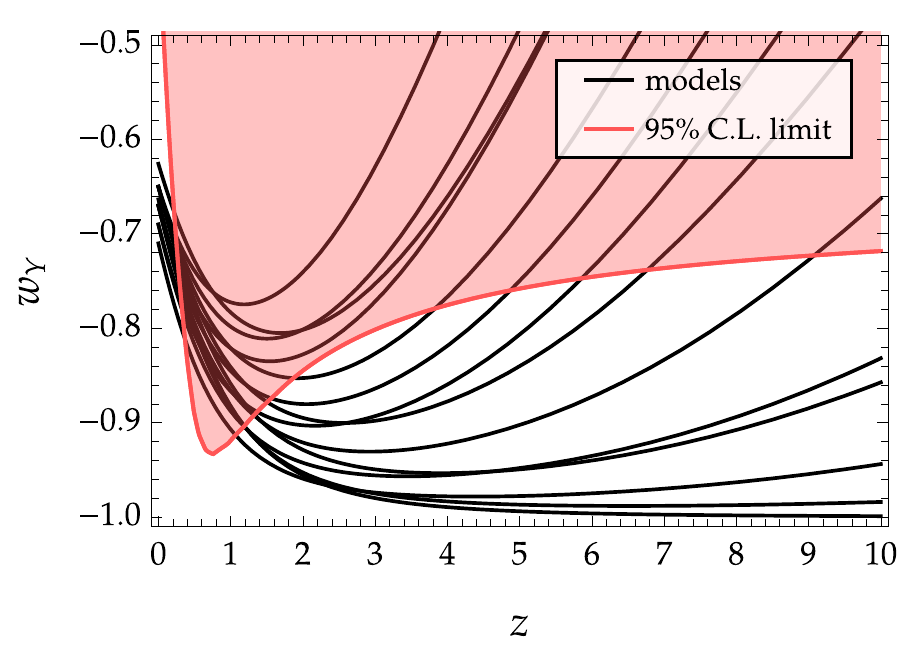}{0.99}
\end{minipage}
\caption{{\it Left.} The 95\%CL
      upper limit on $w_Y(z) = w_0 + w_a z/(1+z)$ based on SNe Ia, CMB and
      BAO data. Following~\cite{Agrawal:2018own}, the limit is determined from Fig.~5 in~\cite{DiValentino:2020evt}
      by finding the values of ($w_0,w_a$) all along the 95\%CL
      contour, plotting all $w_Y(z)$, and finding the upper 
      hull. {\it Right.} A comparison between the 95\% CL upper limit derived
      in the left panel and various predictions for the Salam-Sezgin-Cveti\v c-Gibbons-Pope model. 
\label{fig:1}}
\end{figure*}

A second constraint discussed by MSW pertains
to the equation of state for dark energy as a function of redshift,
$w_Y (z)$. Before proceeding, we pause to note that it is nearly impossible to
constrain a general history of $w_Y(z)$. This is because the dark energy density, which regulates $H(z)$, is
given by an integral over $w_Y(z)$, and hence length scales and the
growth factor involve a further integration over functions of
$H(z)$. Several parametrizations for $w_Y(z)$ have been proposed; see
e.g.~\cite{Efstathiou:1999tm,Chevallier:2000qy,Weller:2001gf,Linder:2002et,Jassal:2004ej}. It has become conventional to phrase  constraints on $w_Y(z)$ in terms
of a linear evolution model, $w_Y(z)= w_0 + w_a \, z/(1+z)$~\cite{Chevallier:2000qy,Linder:2002et}. Indeed, MSW adopt the constraint on $w_Y(z)$ derived in~\cite{Agrawal:2018own} on the
basis of the linear evolution model and the best fit parameters of supernovae type Ia (SNe Ia), cosmic
microwave background (CMB), and baryon acoustic oscillation (BAO)
measurements~\cite{Scolnic:2017caz}. More concretely, when 
Planck 2015 CMB measurements are combined with data from the Pantheon
SNe Ia sample and constraints from BAO the best fit parameters are $w_0 = -1.007 \pm 0.089$ and $w_a = -0.222\pm
 0.407$~\cite{Scolnic:2017caz}. Over and above, when SNe Ia and BAO datasets are combined with the most recent Planck 2018 observations the precision
 on the best fit parameter improves, yielding $w_0 = -0.964 \pm 0.077$
 and $w_a =
 -0.25^{+0.30}_{-0.26}$~\cite{DiValentino:2020evt}. However, recent
 observations provided evidence to support the possibility that intrinsic SNe 
 Ia luminosities could either evolve with
 redshift~\cite{Kim:2019npy,Kang:2019azh} (see however~\cite{Rose:2020shp}), or else correlate with the host star formation
 rate or metallicity~\cite{Timmes:2003xx,Rigault:2014kaa,Jones:2018vbn,Rigault:2018ffm}. All in all, the
 effect of the new SNe Ia systematic uncertainties leads to both a shift in
 the peak and a broadening of the marginalized posterior probability
 distributions from the multi-dimensional fit used to determine the dark energy parameters: when Pantheon SNe Ia,
BAO, and Planck 2018 datasets are combined $w_0 =
-0.85^{+0.15}_{-0.21}$ and $-0.52^{+0.57}_{-0.40}$, whereas when JLA
SNe Ia, BAO, and Planck 2018 datasets are combined $w_0 = -0.70\pm
0.19$ and $w_a = -0.91 \pm 0.52$~\cite{DiValentino:2020evt}. In
Fig.~\ref{fig:1} we show a comparison between the predictions for $w_Y(z)$ of the 
models studied in~\cite{Anchordoqui:2019amx}
and the 95\%CL upper limit on $w_Y(z)$ derived
in~\cite{DiValentino:2020evt}, taking into account SNe Ia
systematics. The predictions of the models are partially consistent with the
upper limit.  Moreover, the 95\%CL upper limit does not come
  from a direct observation to which one can associate statistical and
  systematic errors. It is the result of a multidimensional fit that
  depends on priors. One of such priors is the adopted functional form
  of $w(z)$, which is subject to large theoretical
  uncertainties~\cite{Sharma:2020unh,Perkovic:2020mph}.  If one adopts
  another form for $w(z)$, then the 95\%CL contours used to derive the
  limit could change too.\footnote{For additional considerations on the 95\%CL upper limit, see~\cite{Heisenberg:2018yae,Akrami:2018ylq,Heisenberg:2018rdu}.}
All in all,  we conclude that our
cosmological framework remains phenomenologically viable.

In summary, we have shown that the Friedmann-Robertson-Walker-Salam-Sezgin model and its string realization by
Cveti\v c, Gibbons, and Pope remains a well equipped framework to
describe cosmological observations. Besides, for the sake of
completness, it is important to stress that in (\ref{msw}) there is an implicit assumption of  {\it critical} string theory which does not hold for time dependent solutions. Consider for instance the simplest 
time-dependent exact solution of string theory described by the linear dilaton background in string frame, corresponding to a linearly expanding universe and logarithmic dilaton in the Einstein frame~\cite{Antoniadis:1988aa,Antoniadis:1988vi}. The underline (super-)CFT in the world-sheet is a free coordinate with a background charge, implying a {\it positive} central charge deficit for the internal CFT. Using a 6-dimensional $\sigma$-model, this implies a negatively curved internal manifold violating the Ricci-flatness assumption of the metric $g_{mn}^{\rm RF}$, such as in the model we described above. Alternatively, one may use flat {\it compact} coordinates in a higher dimensional space, since positive central charge deficit increases effectively the critical dimension of string theory. Another property shared by the model we studied here is the 
{\it non-uniform} time dependence of the internal space (i.e., 
internal dimensions may have different time dependence). 
Allowing in general different directions/cycles to
have different time dependence, leaves plenty of room still available
for model builders.

We end with an observation: the fading dark matter
hypothesis relieves tensions in $H_0$ measurements but it does not
fully resolve them. String theory provides a plethora of candidates
for long-lived relics that can modify the expansion rate at recombination and thus affect the evolution of $H$ and $w_Y$~\cite{Anchordoqui:2019amx,Anchordoqui:2011nh,Baumann:2016wac,Anchordoqui:2020znj}. A comprehensive study of the full parameter space is beyond the scope of this Letter and will be presented elsewhere.

\acknowledgements{
 The work of L.A.A. and J.F.S. is supported  by the U.S. National Science Foundation (NSF Grant PHY-1620661) and
  the National Aeronautics and Space Administration (NASA Grant
  80NSSC18K0464). The research of I.A. is funded in part by the
  ``Institute Lagrange de Paris'', and in part by a CNRS PICS grant. The work of D.L. is
  supported by the Origins Excellence Cluster.  Any opinions, findings, and
  conclusions or recommendations expressed in this material are those
  of the authors and do not necessarily reflect the views of the NSF
  or NASA.}

\end{document}